\documentclass[conference]{IEEEtran}
\ifCLASSINFOpdf
\else
\fi
\usepackage{amsmath,dsfont,bbm,epsfig,amssymb,amsfonts,amstext,verbatim,amsopn,cite}
\usepackage{subfigure,multirow,multicol,lipsum,xfrac}
\usepackage{amsthm}
\usepackage{mathtools,amsthm}
\usepackage{perpage}
\usepackage{balance}
\usepackage{url}
\usepackage{amsfonts}
\usepackage{epsfig}
\usepackage[font={small}]{caption}
\usepackage{psfrag}	
\usepackage{etoolbox}
\usepackage{algorithmicx}
\usepackage[Algorithm,ruled]{algorithm}
\usepackage{algpseudocode}
\usepackage{pifont}
\usepackage[utf8]{inputenc}
\usepackage[T1]{fontenc}  
\usepackage[nolist]{acronym}
\usepackage{textcomp}
\usepackage{paralist}
\usepackage{enumitem}
\usepackage{bbm}
\usepackage[process=auto]{pstool}
\usepackage{tikz,pgfplots}
\usetikzlibrary{shapes,arrows}
\usetikzlibrary{fit}
\tikzset{%
  highlight/.style={rectangle,rounded corners,fill=red!15,draw,fill opacity=0.5,thick,inner sep=0pt}
}

\newcommand{\setX}{\mathbb{X}}
\newcommand{\setM}{\mathbb{M}}
\newcommand{\setA}{\mathbb{A}}

\newcommand{\setR}{\mathbb{R}}

\newcommand{\setF}{\mathbb{F}}

\newcommand{\setI}{\mathbb{I}}

\DeclareMathOperator*{\argmin}{argmin}

\newcommand{\rmp}{\mathrm{p}}

\newcommand{\rmg}{\mathrm{g}}

\newcommand{\maa}{\mathcal{A}}

\newcommand{\mac}{\mathcal{C}}

\newcommand{\bvv}{\mathbf{v}}

\newcommand{\bx}{{\boldsymbol{x}}}

\newcommand{\yy}{\mathrm{y}}

\newcommand{\zz}{\mathrm{z}}

\newcommand{\set}[1]{\left\lbrace#1\right\rbrace}
\newcommand{\brc}[1]{\left( #1 \right)}
\newcommand{\dbc}[1]{\left[ #1 \right]}

\newcommand{\sel}{\mathrm{Sel}}

\newcommand{\bz}{{\boldsymbol{z}}}
\newcommand{\ba}{{\mathbf{a}}}

\newcommand{\bu}{{\mathbf{u}}}

\newcommand{\bc}{{\mathbf{c}}}

\newcommand{\baa}{{\mathbf{a}}}

\newcommand{\dif}{\mathrm{d}}

\newcommand{\by}{{\boldsymbol{y}}}

\newcommand{\trp}{\mathsf{T}}

\newcommand{\mA}{\mathbf{A}}
\newcommand{\mW}{\mathbf{W}}

\newcommand{\mP}{\mathbf{P}}
\newcommand{\mI}{\mathbf{I}}

\newcommand{\mE}{\mathbf{E}}

\newcommand{\mQ}{\mathbf{Q}}

\newcommand{\mU}{\mathbf{U}}

\newcommand{\mY}{\mathbf{Y}}

\newcommand{\mF}{\mathbf{F}}

\newcommand{\E}{\mathbb{E}\hspace{.5mm}}

\newcommand{\norm}[1]{\lVert #1 \rVert}

\newcommand{\abs}[1]{\lvert #1 \rvert}

\newtheoremstyle{mystyle}
  {}
  {}
  {}
  {}
  {\bfseries}
  {:}
  { }
  {}
\theoremstyle{mystyle}

\algnewcommand\algorithmicLet{\textbf{Let}}
\algnewcommand\Let{\item[\algorithmicLet]}
\algnewcommand\algorithmicSet{\textbf{Set}}
\algnewcommand\Set{\item[\algorithmicSet]}

\algnewcommand\algorithmicInitiate{\textbf{Initiate}}
\algnewcommand\Initiate{\item[\algorithmicInitiate]}
\algnewcommand\algorithmicStart{\textbf{Begin}}
\algnewcommand\Begin{\item[\algorithmicStart]}
\algnewcommand\algorithmicEnd{\textbf{End}}
\algnewcommand\End{\item[\algorithmicEnd]}

\algnewcommand\algorithmicOutP{\textbf{Output:}}
\algnewcommand\Out{\item[\algorithmicOutP]}

\algnewcommand\algorithmicInP{\textbf{Input:}}
\algnewcommand\In{\item[\algorithmicInP]}

\newcounter{bar}

\begin{document}
\title{Oversampled Adaptive Sensing via\vspace*{1mm}\\ a Predefined Codebook}

\author{
\IEEEauthorblockN{
Ali Bereyhi,
Saba Asaad, and
Ralf R. M\"uller
}
\IEEEauthorblockA{
Friedrich-Alexander Universit\"at Erlangen-N\"urnberg, Germany\\
\{ali.bereyhi, saba.asaad, ralf.r.mueller\}@fau.de
\thanks{This work has been presented in 2021 IEEE International Symposium on Joint Communications \& Sensing (JC\&S). The link to the final version in the proceedings will be available later.}
}
}


\IEEEoverridecommandlockouts

\maketitle

\begin{acronym}
\acro{oas}[OAS]{oversampled adaptive sensing}
\acro{awgn}[AWGN]{additive white Gaussian noise}
\acro{iid}[i.i.d.]{independent and identically~dis-tributed}
\acro{rhs}[r.h.s.]{right hand side}
\acro{lhs}[l.h.s.]{left hand side}
\acro{wrt}[w.r.t.]{with respect to}
\acro{mse}[MSE]{mean squared error}
\acro{mmse}[MMSE]{minimum mean squared error}
\acro{snr}[SNR]{signal-to-noise ratio}
\acro{sinr}[SINR]{signal to interference and noise ratio}
\acro{mf}[MF]{match filtering}
\end{acronym}

\begin{abstract}
Oversampled adaptive sensing (OAS) is a Bayesian framework recently proposed for effective sensing of \textit{structured} signals in a \textit{time-limited} setting. In contrast to the conventional \textit{blind} oversampling, OAS uses the prior information on the signal to construct posterior beliefs sequentially. These beliefs help in constructive oversampling which iteratively evolves through a sequence of time sub-frames. 

The initial studies of OAS consider the idealistic assumption of full control on sensing coefficients which is not feasible in many applications. In this work, we extend the initial investigations on OAS to more realistic settings in which the sensing coefficients are selected from a predefined set of possible choices, referred to as the \textit{codebook}. We extend the OAS framework to these settings and compare its performance with classical non-adaptive approaches.
\end{abstract}


\IEEEpeerreviewmaketitle

\section{Introduction}
\label{sec:intro}
Consider the following classical problem which raises in several sensing scenarios: A set of signal samples $x_1,\ldots,x_N$ are to be sensed in a noisy environment via $K$ sensors within a limited time frame. Each sensor is tunable and can observe various linear combinations of signal samples. The ultimate goal is to collect some observations from which the signal samples are recovered with minimum distortion. There are in general two approaches to address this goal:
\begin{enumerate}
	\item The undersampled \textit{non-adaptive} approach, in which the sensors are tuned once at the beginning of the time frame and kept fixed for the whole duration. In this case, $K$ high quality observations, i.e., observations with high \ac{snr}, are collected. These are then given to an estimator to recover the signal samples.
	\item The \ac{oas} framework \cite{muller2018oversampled,muller2018randomoversampled,bereyhi2019oas}, in which the given time frame is divided into multiple subframes. In each subframe, $K$ observations are collected. These observations, along with those collected in previous subframes, are used to decide for a new tuning strategy in the next subframe. Clearly in this case, a larger number of observations are collected. This increase in the number of observations is obtained at the expense of lower quality for each particular observation\footnote{Note that since subframes are shorter than the whole time frame, the noise variance is higher. This results in lower \ac{snr} for a particular observation. This point is illustrated in the Section~\ref{sec:SysMod}.}.
\end{enumerate}

Despite initial studies on adaptive approaches for signal sensing, e.g., \cite{haupt2009compressive,haupt2009adaptive,malloy2014near}, it was commonly believed in the literature that adaptation does not result in a significant performance enhancement, assuming that the trade-off between the \textit{quality} and \textit{quantity} of the observations leads to no performance gain. In \cite{muller2018oversampled}, this myth was shown to be wrong.  \ac{oas} illustrates this trade-off between the number of observations and their quality as follows: For cases with some prior information on the signal samples, e.g., sparse recovery, the adaptation can result on a significant enhancement. Nevertheless, in lack of prior information, the conventional belief holds. 

The superiority of \ac{oas} framework is intuitively understood as follows: In presence of prior information, several samples\footnote{For instance, zero samples in the example of sparse recovery.} are effectively recovered in the primary subframes. By omitting those samples, the space of unknowns shrinks, and hence the recovery can be performed effectively.  

\subsection{Contributions}
The initial studies of \ac{oas} consider no restrictions on the sensor tuning. This is an idealistic assumption which is not necessarily satisfied in practice; see for instance the example given in Section~\ref{sec:sys2}. It is hence more realistic to assume that the tuning of sensors is controlled by a predefined codebook which includes available choices of tuning. Given this new restriction, this study develops an \ac{oas} algorithm which collects observations in each subframe using a predefined codebook. The performance of the proposed algorithm is investigated in various respects, and the impacts of limited codebook size are clarified through several experiments.

\subsection{Motivation and Applications}
\label{sec:sys2}
To give an intuition on sensing via a predefined codebook, let us consider the following example: A signal is sampled~via a preinstalled sensor network within a limited time frame.~At a given time, only a subset of sensors are active in~the~network. To apply \ac{oas}, the time frame is divided into several subframes, and the active sensors are altered in each subframe. Clearly, the sensing procedure in this setting is restricted by the preinstalled network and cannot be changed arbitrarily. 

Similar to the given example, in several practical scenarios, sensing is performed by means of a preinstalled setting whose possible configurations for observing a signal is restricted by  a codebook. Examples of such scenarios are found in radar and positioning systems. In these applications, it is often the case that prior information is available on the target signal; for instance, in many radar systems, the observed signal is known to be sparse. The main motivation of this study is to extend the scope of \ac{oas} to these applications.


\subsection{Notation}
Scalars, vectors and matrices are shown with non-bold, bold lower case and bold upper case letters, respectively. $\mI_K$ and $\boldsymbol{0}_{K\times N}$ are the $K \times K$ identity matrix and $K \times N$~all-zero matrix, respectively. $\mA^{\trp}$ denotes the transpose of $\mA$. The set of real numbers is shown by $\setR$. We use the shortened~notation $[N]$ to represent $\set{1, \ldots , N}$.

\section{Problem Formulation}
\label{sec:SysMod}
Consider $\bx\in\setR^N$ containing $N$ signal samples. We postulate\footnote{This postulation is not necessarily \textit{matched} to the true distribution.} that the samples are \ac{iid} with prior distribution $q\brc{x}$. These~samples are observed via $K$ sensors within a restricted time interval of duration $T$. A particular sensor $k \in \dbc{K}$ observes a noisy linear combination of signal samples, i.e., 
\begin{align}
y_k = \left. \baa_k^\trp  \right. \bx + z_k \label{eq:y_ell}
\end{align}
for additive noise $z_k$ and the \textit{vector of coefficients} $\baa_k\in \setR^N$:
\begin{enumerate}
	\item The noise power is reversely proportional to the observation time: Assume that sensor $k$ operates for a time frame of length $t \leq T$. Then, $z_k$ is assumed to be zero-mean Gaussian with variance $\sigma^2\brc{t} = {\sigma^2_0}/ {t}$ for some $\sigma^2_0 \geq 0$.
	\item The coefficient vectors $\baa_k$ for $k\in\dbc{K}$ are selected from a predefined \textit{codebook} $\mac = \set{ \bc_1, \ldots,\bc_S }$ where $S\geq K$ and $\bc_s \in\setR^N$ for $s\in \dbc{S}$. This means that 
	\begin{align}
		\set{ \baa_1, \ldots,\baa_K } \subseteq \mac = \set{ \bc_1, \ldots,\bc_S }.
	\end{align}
\end{enumerate}

There are two key points in the sensing model \eqref{eq:y_ell} which deviate from the classical models: 
\begin{enumerate}
	\item The sensing quality is \textit{time-dependent}.
	\item The sensing is performed using a predefined codebook.
\end{enumerate}

The time-dependent model follows the fact that the \ac{snr} of a particular observation linearly scales with the duration of observation\footnote{See \cite{muller2018oversampled} for detailed illustrations.}. This is a typical assumption in \ac{oas}. The latter point is the key difference of the considered system model, to the typical settings considered for \ac{oas}.

%

\section{OAS Framework}
In a nutshell, the \ac{oas} framework can be represented via the following steps:
\begin{enumerate}[label=(\alph*)]
\item The time frame is divided into $M$ \textit{subframes}.
\item In subframe $m\in\dbc{M}$, the sensors observe
\begin{align}
\by_m = \left. \mA_m \right. \bx + \bz_m, \label{eq:y_m}
\end{align}
where $\bz_m\sim \mathcal{N} \brc{ \boldsymbol{0}, \sigma^2_{\rm s} \mI_K }$ with $\sigma^2_{\rm s} = M\sigma^2\brc{T}$, and
\begin{align}
	\mA_m = \dbc{ \baa_1\brc{m}, \ldots, \baa_K\brc{m} }^\trp
\end{align}
with $\set{\baa_1\brc{m}, \ldots, \baa_K\brc{m} } \subseteq \mac$ containing the vectors selected in subframe $m$. 
\item A processing unit collects the observations up to subframe $m$ in a matrix of \textit{stacked observations}
\begin{align}
\mY_m \coloneqq \dbc{ \by_1, \ldots, \by_m}.
\end{align}
It then uses a \textit{Bayesian estimator} to calculate some~\textit{soft information}. In general, the soft information is written as
\begin{align}
\hat{\bx}_m  = \E \set{\bx | \mY_m , \setA_m},
\end{align}
where $\setA_m$ denotes the collection of sensing matrices up to subframe $m$, i.e.,
\begin{align}
\setA_m = \set{ \mA_1, \ldots,\mA_m },
\end{align}
and the expectation is taken with respect to the posterior distribution $\rmp\brc{\bx \vert \mY_m , \setA_m}$ which is calculated from the likelihood using the postulated prior distribution $q\brc{x}$.
\item The processing unit specifies the sensing matrix of subframe $m+1$ using the soft information. This means that
\begin{align}
\mA_{m+1} = f_{\rm Adp} \brc{\hat{\bx}_m}	
\end{align}
for some adaptation function $f_{\rm Adp} \brc\cdot$.
\end{enumerate}

\subsection{Performance Characterization}
The conventional metric which quantifies the recovery performance is the \textit{average distortion}: Let $\Delta\brc{\cdot;\cdot} : \setR \times \setR \mapsto \setR_0^+$ be a distortion function\footnote{For instance, the Euclidean distance.}. The distortion between two vectors $\bx,\hat\bx \in \setR^N$ with respect to $\Delta\brc{\cdot;\cdot}$ is determined as
\begin{align}
\Delta\brc{\bx; \hat\bx } = \sum_{n=1}^N \Delta\brc{x_n; \hat{x}_n }.
\end{align}

In an \ac{oas}-based algorithm, the signal samples are finally recovered as ${\bx^\star} = \rmg \brc{\hat{\bx}_M}$ where $\rmg\brc{\cdot}$ is a \textit{decision function}\footnote{For example, a hard or soft thresholding function.} used to recover the signal samples from the soft estimate $\hat{\bx}_M$. Consequently, the average distortion is 
\begin{align}
D = \E \set{\Delta\brc{\bx; \bx^\star } }.
\end{align}

A non-adaptive recovery technique can be seen as an \ac{oas}-based algorithm with a single time subframe. The performance in this case is hence characterized by setting $M=1$.

\section{OAS via a Predefined Codebook}
In settings with no predefined codebook, designing \ac{oas}-based algorithms deals with high degrees of freedom. In fact, sensing matrices are freely constructed in such scenarios. With a predefined codebook, algorithm~design is significantly restricted, since degrees of freedom are limited via the codebook.

In the sequel, we discuss the design via a sample algorithm which performs \ac{oas} via a predefined codebook. The algorithm is given in Algorithm~\ref{alg:OAS}. In this algorithm, 
\begin{itemize}
	\item The selector operator $\sel\brc{\setX}$, for an index set $\setX\subseteq \dbc{N}$ with $L$ indices, returns an $L\times N$ matrix. Setting $\mP =\sel\brc{\setX}$, the $\ell$-th row of $\mP$ is the standard basis of $\setR^N$ whose non-zero entry occurs at the $\ell$-th index of $\setX$, e.g., for $N=4$ and $\setX = \set{1,3}$, we have
	\begin{align}
		\sel\brc{\setX} = \begin{bmatrix}
			1 &0 &0 &0\\
			0 &0 &1 &0
		\end{bmatrix}.
	\end{align}
	\item The posterior distribution $\rmp_0 \brc{u \vert y , \sigma^2}$ is a scalar distribution and is given by
	\begin{align*}
		\rmp_0 \brc{u \vert y , \sigma^2} = \frac{ \displaystyle \exp\set{- \frac{\brc{y - u }^2 }{  2  \sigma^2 } } q\brc{u} }{\displaystyle  \int \exp\set{- \frac{ \brc{y - u }^2 }{2 \sigma^2 }}  q\brc{u} \dif u}.
	\end{align*}
\item $\maa \brc{ \mac, K , \setF, \hat{\bx} }$ is a \textit{selection algorithm} selecting $K$ vectors from the codebook $\mac$ in order to sense signal samples which are indexed by $\setF$ using the current estimate $\hat{\bx}$.
\end{itemize}

\begin{algorithm}[t]
	\caption{OAS via a Predefined Codebook}
	\label{alg:OAS}
	\begin{algorithmic}[0]
		\In $K$, $L$, codebook $\mac$ and postulated prior $q(x_n)$.
		\Initiate For $n\in\dbc{N}$, set
		$d_{n} = +\infty$, $\hat{x}_n=\yy_n =\hat{\sigma}_n^2 =0$ and $\setM_n \brc{0} = \emptyset$.\vspace*{.5mm}
		\For{$m \in \dbc{M}$}\\
		\begin{enumerate}
			\item Sort $d_1, \ldots d_N$ as $d_{i_1^m}, \ldots d_{i_N^m}$ such that
			\begin{align*}
				d_{i_1^m}\geq  \ldots \geq d_{i_N^m}.
			\end{align*}
			\item Set $\setF_m$ by worst-case adaptation as
			\begin{align*}
				\setF_m = \set{ {i_1^m} , {i_L^m} },
			\end{align*}
			and update $\setM_n \brc{m} = \setM_n \brc{m-1} \cup \set{m}$ for $n\in \setF_m$.
			\item Set $\mA_m = \dbc{ \baa_1\brc{m}, \ldots, \baa_K\brc{m} }^\trp$ where
			\begin{align*}
				\set{ \baa_1\brc{m}, \ldots, \baa_K\brc{m} } = \maa \brc{ \mac, K , \setF_m, \hat{\bx} }
			\end{align*}
			\item Sense the samples for duration $T/M$ via $\mA_m$.
			\item Set $\mP_m=\sel \brc{\setF_m}$ and $\mE_m = \sel \brc{\dbc{N}\backslash\setF_m}$. Let $\tilde{\bx}_m = \mE_m \hat{\bx}$, $\mQ_m = \mA_m \mP_m^\trp$ and $\mW_m = \mA_m \mE_m^\trp$, and set
			\begin{align*}
				\mF_m &= \brc{\mQ_m^\trp \mQ_m}^{-1} \mQ_m^\trp
			\end{align*}
			\item Decouple the observations as
			\begin{align*}
				\hat{\by}_m = \mF_m \brc{ \by_m - \mW_m \tilde{\bx}_m}.
			\end{align*}
			\item For $\ell\in\dbc{L}$, update
			\begin{align*}
				\yy_{i_\ell^m} &= \yy_{i_\ell^m} +  \hat{y}_{m,\ell}\\
				\hat{\sigma}_{i_\ell^m}^2 &= \hat{\sigma}_{i_\ell^m}^2 + M \norm{\mathbf{f}_{m,\ell} }^2 \sigma^2
			\end{align*}
			with $\mathbf{f}_{m,\ell}^\trp$ being the $\ell$-th row of $\mF_m$.
			\item For $n\in\setF_m$, update
			\begin{align*}
				\hat{x}_{n} &=  \int u \; \rmp_0 \brc{u  \left\vert \frac{\yy_{n}}{\abs{ \setM_n \brc{m} } },  \frac{ \hat{\sigma}_{n}^2 }{\abs{ \setM_n \brc{m} }^2 } \right.  }  \dif u\\
				{d}_{n} &=  \int \brc{u-\hat{x}_n}^2 \rmp_0 \brc{u  \left\vert \frac{\yy_{n}}{\abs{ \setM_n \brc{m} } },  \frac{ \hat{\sigma}_{n}^2 }{\abs{ \setM_n \brc{m} }^2 } \right.  }  \dif u\\
			\end{align*}
		\end{enumerate}
		\EndFor
	\end{algorithmic}
\end{algorithm}

\subsection{Derivation of Algorithm~\ref{alg:OAS}}
\label{sec:derivations}
In a nutshell, Algorithm~\ref{alg:OAS} follows these steps: It first focuses on a subset of $L\leq K$ signal samples via worst-case adaptation and selects $K$ vectors from the codebook to sense them. Using the fact that recovery of this subset is a determined problem, the algorithm uses linear filtering to decouple the observations. It then cancels out the other $N-L$ signal samples using the estimates obtained in previous subframes and estimates the $L$ samples via a scalar Bayesian estimator assuming that the residual terms are Gaussian. 

To derive the update rules given in the algorithm, consider subframe $m-1$ at which the signal samples are estimated as $\hat{x}_1, \ldots,\hat{x}_N$ and \textit{posterior distortions} $d_1, \ldots,d_N$ are calculated for the signal samples. The posterior distortion $d_n$ for the $n$-th signal sample is determined using $y_n$, which is a \textit{statistic} on $x_n$, and \textit{estimated noise variance} $\hat{\sigma}^2_n$ as
\begin{align}
{d}_{n} &=  \int \brc{u-\hat{x}_n}^2 \rmp_0 \brc{u  \vert y_n, \hat{\sigma}^2_n }  \dif u.
\end{align}
The statistic and estimated noise variance are calculated from $\mY_{m-1}$ and $\setA_{m-1}$ in a way that will be clarified in the sequel.
	
 For $L\leq K$, the algorithm finds $L$ samples whose posterior distortions are maximum. The indices of these samples are collected in $\setF_m \subseteq \dbc{N}$. A selection algorithm $\maa \brc{ \mac, K , \setF_m, \hat{\bx} }$ is then used to select $K$ vectors from the codebook $\mac$. At this point, we consider a generic selection algorithm, and leave the discussions on the design of $\maa \brc{ \mac, K , \setF_m, \hat{\bx} }$ for later. 

Let $\mA_m$ be the collection of vectors selected in subframe $m$. The observations are hence given by \eqref{eq:y_m} which read 
\begin{subequations}
	\begin{align}
	\by_m &= \mA_m \bx + \bz_m\\
	&=\brc{\mA_m \mP_m^\trp \mP_m +  \mA_m \mE_m^\trp \mE_m} \bx + \bz_m
\end{align}
\end{subequations}
where $\mP_m = \sel\brc{\setF_m}$ and $\mE_m = \sel\brc{\dbc{N} \backslash \setF_m}$. Defining the matrices $\mQ_m = \mA_m \mP_m^\trp$ and $\mW_m = \mA_m \mE_m^\trp$, we can write
\begin{align}
	\by_m =\mQ_m \breve{\bx}_m +  \mW_m \bar{\bx}_m + \bz_m
\end{align}
where  $\breve{\bx}_m = \mP_m \bx$ contains the $L$  samples selected by $\setF_m$ and $\bar{\bx}_m = \mE_m \bx$ consists of the remaining $N-L$ signal samples. Since the entries of $\bar{\bx}_m$ are not estimated in this subframe, we approximate them by the estimates of the previous subframe, i.e., $\bar{\bx}_m \approx \tilde{\bx}_m = \mE_m \hat{\bx}$. We hence can write
\begin{align}
	\by_m - \mW_m \tilde{\bx}_m   \approx \mQ_m \breve{\bx}_m + \bz_m. \label{eq:approx}
\end{align}

Noting that $\mQ_m \in \setR^{K \times L}$ with $K \geq L$, we could conclude that for a proper selection of codewords $\mQ_m^\trp \mQ_m$ is full-rank, and thus, we can calculate the pseudo-inverse of $\mQ_m$ as
\begin{align}
	\mF_m &= \brc{\mQ_m^\trp \mQ_m}^{-1} \mQ_m^\trp.
\end{align}
Consequently, the observations can be decoupled into $L$ samples with additive noise as 
\begin{align}
	\hat{\by}_m &= \mF_m \brc{\by_m - \mW_m \tilde{\bx}_m}\\
	  & \approx \breve{\bx}_m + \mF_m \bz_m.
\end{align}

We now consider this \textit{mismatched assumption} that $\mF_m \bz_m$ is a Gaussian vector with \textit{independent entries}\footnote{Such an assumption is asymptotically correct for large random codebooks with \ac{iid} entries.}. As the result, the $\ell$-th entry of $\hat{\by}_m$ can be written as
\begin{align}
	\hat{y}_{m,\ell} \approx x_{i_\ell^m} + \hat{z}_{m,\ell} 
\end{align}
where $i_\ell^m$ denotes the $\ell$-th entry of $\setF_m$, and \textit{decoupled noise} $\hat{z}_{m,\ell} $ is zero-mean Gaussian with variance $M \norm{\mathbf{f}_{m,\ell} }^2 \sigma^2$. Here, $\mathbf{f}_{m,\ell}^\trp$ denotes the $\ell$-th row of $\mF_m$.

We now use $\hat{y}_{m,\ell}$ to update the \textit{statistic} on $x_{i_\ell^m}$ by adding the decoupled observation corresponding to $x_{i_\ell^m}$ to the current statistic: Let 	$\yy_{i_\ell^m} \brc{m-1}$ denote the statistic on $x_{i_\ell^m}$ collected in subframes $1, \ldots,m-1$. We update this parameter as
\begin{align}
	\yy_{i_\ell^m} \brc{m}&= \yy_{i_\ell^m} \brc{m-1} +  \hat{y}_{m,\ell}.
\end{align}

Let $\setM_n\brc{m}\subseteq \dbc{m}$ collects all subframes at which $n \in \setF_m$, i.e., $x_n$ is selected to be sensed in these subframes. Moreover, assume that $\hat{\sigma}_n^2\brc{m}$ denotes the sum of all decoupled noise variances in $\setM_n\brc{m}$ which correspond to index $n$. After updating our statistic on $x_{i_\ell^m}$, we have
\begin{align}
	\yy_{i_\ell^m} \brc{m} \approx \abs{\setM_n\brc{m}} x_{i_\ell^m} + \zz_{i_\ell^m} \brc{m}
\end{align}
where $\zz_{i_\ell^m} \brc{m}$ is  given by
\begin{align}
	\zz_{i_\ell^m} \brc{m} = \zz_{i_\ell^m} \brc{m-1} + \hat{z}_{i_\ell^m} \brc{m}.
\end{align}
Hence, the \textit{estimated noise variance} is updated as
\begin{align}
	\hat{\sigma}_{i_\ell^m}^2 \brc{m} = \hat{\sigma}_{i_\ell^m}^2 \brc{m-1} + M \norm{\mathbf{f}_{m,\ell} }^2 \sigma^2.
\end{align}
Using the Bayesian estimator, we finally update the estimate $\hat{x}_n$ and the posterior distortion $d_n$ via the updated statistics $\yy_n$ for $n\in \setF_m$. This concludes Algorithm~\ref{alg:OAS}.

\section{Example of Sparse Recovery}
Initial investigations demonstrate that \ac{oas} can significantly outperform classical sparse recovery approaches. This is due to the prior information on the sparsity of signal samples. In this section, we employ our algorithm to address this example.

For a sparse signal, the prior distribution of the samples is
\begin{align}
	q\brc{x} = \brc{1-\rho} \delta\brc{x} + \rho \phi \brc{x}
\end{align}
for sparsity factor $\rho \in \dbc{0,1}$ and a distribution $\phi \brc{x}$ satisfying
\begin{align}
	\int_{0^-}^{0^+} \phi \brc{x} = 0.
\end{align}
The model indicates that samples are zero with probability $1-\rho$ and are drawn from distribution $\phi \brc{x}$ with probability $\rho$. In the sequel, we consider an \textit{\ac{iid} sparse-Gaussian} signal for which the samples are \ac{iid} and 
\begin{align}
 \phi \brc{x} = \frac{1}{\sqrt{2\pi}} \exp\set{ - \frac{x^2}{2}}.
\end{align}

Considering the system model, the main task in this example is to recover the samples of a given sparse signal using the observations made within a time frame of length $T$ via the $K$ active sensors. This task can be addressed in two ways:
\begin{itemize}
	\item Invoking the theory of compressive sensing, one~can~select a sensing matrix from the codebook, and collect $K$ observations. A classical sparse recovery algorithm, e.g., LASSO, is then used to recover the signal samples. 
	\item An alternative approach is to use an \ac{oas} algorithm, e.g., Algorithm~\ref{alg:OAS}, to adaptively estimate the signal samples.
\end{itemize}
We investigate these two approaches in the sequel.

\subsection{Sparse Recovery via OAS}
For a sparse-Gaussian prior distribution, the Bayesian estimation in Algorithm~\ref{alg:OAS} reduces to
\begin{align*}
	\hat{x}_{n} &= G \brc{ \frac{\yy_{n}}{\abs{ \setM_n \brc{m} } }  \left\vert  \frac{ \hat{\sigma}_{n}^2 }{\abs{ \setM_n \brc{m} }^2 } \right. }, \\
	d_{n} &= E \brc{ \frac{\yy_{n}}{\abs{ \setM_n \brc{m} } }  \left\vert  \frac{ \hat{\sigma}_{n}^2 }{\abs{ \setM_n \brc{m} }^2 } \right. }.
\end{align*}
where $G\brc{ y \vert \sigma^2 }$ and $E\brc{ y \vert \sigma^2 }$ are given by
\begin{subequations}
	\begin{align}
		G\brc{ y \vert \sigma^2 } &=   \frac{I_1 \brc{ y \vert \sigma^2 } y}{J \brc{ y \vert \sigma^2 } }\\
		E\brc{ y \vert \sigma^2 } &=   \frac{I_1 \brc{ y \vert \sigma^2 }  }{J \brc{ y \vert \sigma^2 } } \brc{ \sigma^2 +  \frac{I_0 \brc{ y \vert \sigma^2 }  }{J \brc{ y \vert \sigma^2 } } y^2
		}
	\end{align}
\end{subequations}
with $I_0 \brc{ y \vert \sigma^2 }$, $I_1 \brc{ y \vert \sigma^2 }$ and $J \brc{ y \vert \sigma^2 }$ being
\begin{subequations}
	\begin{align}
		I_0 \brc{ y \vert \sigma^2 } &=  \brc{1-\rho} \sqrt{1+\sigma^2} \exp\set{- \frac{y^2}{2{\sigma^2}}}\\
		I_1 \brc{ y \vert \sigma^2 } &=  \rho \; \sigma \exp\set{- \frac{y^2}{2\brc{1+\sigma^2}}}\\
		J \brc{ y \vert \sigma^2 } &= \brc{1+\sigma^2} \dbc{	I_0 \brc{ y \vert \sigma^2 } +	I_1 \brc{ y \vert \sigma^2 }  }.
	\end{align}
\end{subequations}

\section{Numerical Experiments}
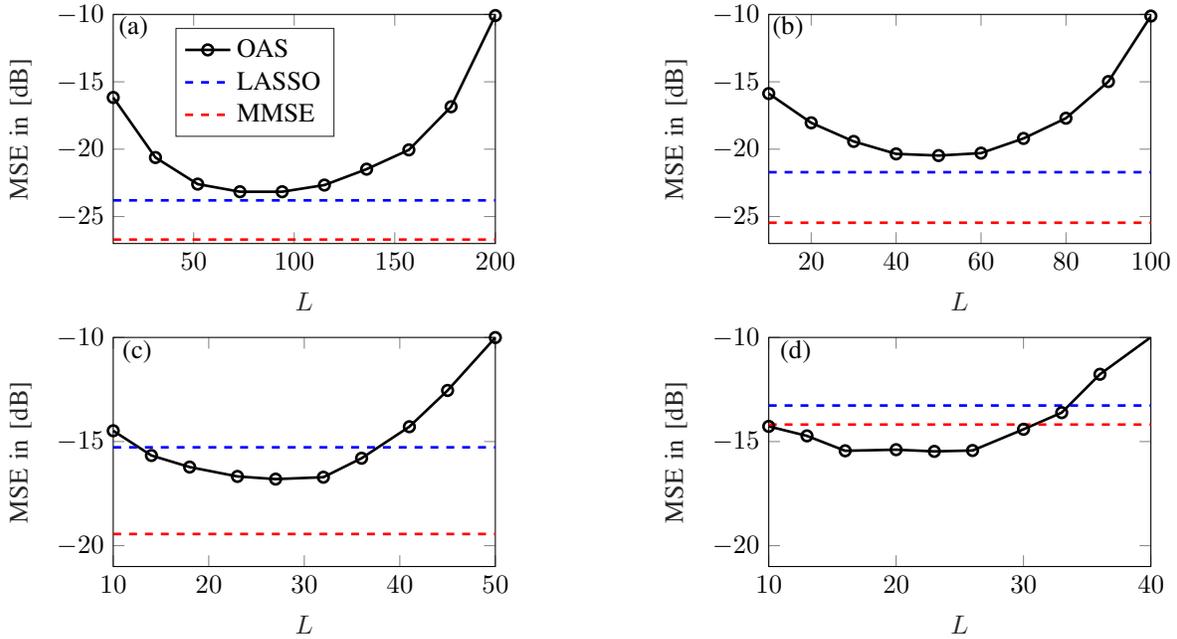
\begin{figure*}[t]
	\centering
%
%
\begin{tikzpicture}

\begin{axis}[%
width=2in,
height=1.2in,
at={(1.013in,2.692in)},
scale only axis,
xmin=10,
xmax=200,
xlabel style={font=\color{white!15!black}},
xlabel={$L$},
ymin=-27,
ymax=-10,
ylabel style={font=\color{white!15!black}},
ylabel={MSE in [dB]},
axis background/.style={fill=white},
legend style={at={(axis cs: 120,-11)},legend cell align=left, align=left, draw=white!15!black}
]

\addplot [color=black, line width=1.0pt, mark=o, mark options={solid, black}]
  table[row sep=crcr]{%
10	-16.1620415725284\\
31	-20.6189391059571\\
52	-22.5923816066717\\
73	-23.1568445670804\\
94	-23.1575216819078\\
115	-22.6629958478077\\
136	-21.4823683920346\\
157	-20.051049524429\\
178	-16.851122113676\\
200	-10.0830266732769\\
};
\addlegendentry{OAS}

\addplot [color=blue, line width=1.0pt, dashed]
  table[row sep=crcr]{%
10	-23.7988424961844\\
31	-23.7988424961844\\
52	-23.7988424961844\\
73	-23.7988424961844\\
94	-23.7988424961844\\
115	-23.7988424961844\\
136	-23.7988424961844\\
157	-23.7988424961844\\
178	-23.7988424961844\\
200	-23.7988424961844\\
};
\addlegendentry{LASSO}

\addplot [color=red, line width=1.0pt, dashed]
  table[row sep=crcr]{%
10	-26.7090331369231\\
31	-26.7090331369231\\
52	-26.7090331369231\\
73	-26.7090331369231\\
94	-26.7090331369231\\
115	-26.7090331369231\\
136	-26.7090331369231\\
157	-26.7090331369231\\
178	-26.7090331369231\\
200	-26.7090331369231\\
};
\addlegendentry{MMSE}

\node at (axis cs:20,.-11) {(a)};

\end{axis}

\begin{axis}[%
width=2in,
height=1.2in,
at={(4.444in,2.692in)},
scale only axis,
xmin=10,
xmax=100,
xlabel style={font=\color{white!15!black}},
xlabel={$L$},
ymin=-27,
ymax=-10,
ylabel style={font=\color{white!15!black}},
ylabel={MSE in [dB]},
axis background/.style={fill=white},
legend style={legend cell align=left, align=left, draw=white!15!black}
]
\addplot [color=black, line width=1.0pt, mark=o, mark options={solid, black}, forget plot]
  table[row sep=crcr]{%
10	-15.8736113776941\\
20	-18.0424106202111\\
30	-19.4248801790832\\
40	-20.3486392383628\\
50	-20.4732877835404\\
60	-20.2917631073408\\
70	-19.2027159335496\\
80	-17.7012595749485\\
90	-14.9816095570254\\
100	-10.121953790808\\
};

\addplot [color=blue, line width=1.0pt, dashed, forget plot]
  table[row sep=crcr]{%
10	-21.7087863143831\\
20	-21.7087863143831\\
30	-21.7087863143831\\
40	-21.7087863143831\\
50	-21.7087863143831\\
60	-21.7087863143831\\
70	-21.7087863143831\\
80	-21.7087863143831\\
90	-21.7087863143831\\
100	-21.7087863143831\\
};

\addplot [color=red, line width=1.0pt, dashed, forget plot]
  table[row sep=crcr]{%
10	-25.4574314892307\\
20	-25.4574314892307\\
30	-25.4574314892307\\
40	-25.4574314892307\\
50	-25.4574314892307\\
60	-25.4574314892307\\
70	-25.4574314892307\\
80	-25.4574314892307\\
90	-25.4574314892307\\
100	-25.4574314892307\\
};
\node at (axis cs:14.5,.-11) {(b)};

\end{axis}

\begin{axis}[%
width=2in,
height=1.2in,
at={(1.013in,1in)},
scale only axis,
xmin=10,
xmax=50,
xlabel style={font=\color{white!15!black}},
xlabel={$L$},
ymin=-21,
ymax=-10,
ylabel style={font=\color{white!15!black}},
ylabel={MSE in [dB]},
axis background/.style={fill=white},
legend style={legend cell align=left, align=left, draw=white!15!black}
]
\addplot [color=black,  line width=1.0pt, mark=o, mark options={solid, black}, forget plot]
  table[row sep=crcr]{%
10	-14.4803968823365\\
14	-15.668516167458\\
18	-16.2238785767465\\
23	-16.6736306873151\\
27	-16.8019157731149\\
32	-16.7113100299524\\
36	-15.7995116891045\\
41	-14.2864405384455\\
45	-12.5452033894275\\
50	-10.005328458206\\
};

\addplot [color=blue, line width=1.0pt, dashed, forget plot]
  table[row sep=crcr]{%
10	-15.2756881199302\\
14	-15.2756881199302\\
18	-15.2756881199302\\
23	-15.2756881199302\\
27	-15.2756881199302\\
32	-15.2756881199302\\
36	-15.2756881199302\\
41	-15.2756881199302\\
45	-15.2756881199302\\
50	-15.2756881199302\\
};

\addplot [color=red, line width=1.0pt, dashed, forget plot]
  table[row sep=crcr]{%
10	-19.4385388842104\\
14	-19.4385388842104\\
18	-19.4385388842104\\
23	-19.4385388842104\\
27	-19.4385388842104\\
32	-19.4385388842104\\
36	-19.4385388842104\\
41	-19.4385388842104\\
45	-19.4385388842104\\
50	-19.4385388842104\\
};

\node at (axis cs:12.5,.-10.7) {(c)};

\end{axis}

\begin{axis}[%
width=2in,
height=1.2in,
at={(4.444in,1in)},
scale only axis,
xmin=10,
xmax=40,
xlabel style={font=\color{white!15!black}},
xlabel={$L$},
ymin=-21,
ymax=-10,
ylabel style={font=\color{white!15!black}},
ylabel={MSE in [dB]},
axis background/.style={fill=white},
legend style={legend cell align=left, align=left, draw=white!15!black}
]
\addplot [color=black, line width=1.0pt, mark=o, mark options={solid, black}, forget plot]
  table[row sep=crcr]{%
10	-14.2705227085893\\
13	-14.7269245936189\\
16	-15.4427483931612\\
20	-15.389294537565\\
23	-15.4708642817175\\
26	-15.4290223202941\\
30	-14.410390168543\\
33	-13.6105786328416\\
36	-11.7685903113014\\
40	-9.98811317087531\\
};

\addplot [color=blue, line width=1.0pt, dashed, forget plot]
  table[row sep=crcr]{%
10	-13.2684000962942\\
13	-13.2684000962942\\
16	-13.2684000962942\\
20	-13.2684000962942\\
23	-13.2684000962942\\
26	-13.2684000962942\\
30	-13.2684000962942\\
33	-13.2684000962942\\
36	-13.2684000962942\\
40	-13.2684000962942\\
};

\addplot [color=red, line width=1.0pt, dashed, forget plot]
  table[row sep=crcr]{%
10	-14.1837317839488\\
13	-14.1837317839488\\
16	-14.1837317839488\\
20	-14.1837317839488\\
23	-14.1837317839488\\
26	-14.1837317839488\\
30	-14.1837317839488\\
33	-14.1837317839488\\
36	-14.1837317839488\\
40	-14.1837317839488\\
};

\node at (axis cs:12.1,.-10.7) {(d)};

\end{axis}
\end{tikzpicture}%
	\caption{MSE versus $L$ for various compression rates: (a) $R=1$, (b) $R=2$, (c) $R=4$, and (d) $R=5$. Here, the time frame is divided to $M=80$ subframes and $S=1000$.}
	\label{fig:mse_vs_rate}
\end{figure*}

For numerical investigations, we consider a sparse-Gaussian signal with $N=200$ samples whose sparsity factor is $\rho = 0.1$. The noise variance for an interval of length $T$ is $\sigma^2 = 0.01$. This means that by dividing the time frame into $M$ subframes, the noise variance in each subframe is $0.01 M$. The compression rate is further defined as
	$R = {N}/{K}$.

The codebook is generated randomly from an \ac{iid} Gaussian ensemble. This means that the vectors in $\mac$ are independent and have \ac{iid} zero-mean Gaussian entries with variance $1/\sqrt{K}$.

As benchmarks, we consider a \textit{classic compressive sensing setting} in which the $K$ observations are collected using $K$ randomly selected vectors. We evaluate the performance for two recovery algorithms:
\begin{inparaenum}
	\item \textit{LASSO} in which the samples are recovered via regularized $\ell_1$-norm minimization, and
	\item \textit{\Ac{mmse}} estimator in which the samples are recovered via the \textit{optimal} Bayesian estimator.
\end{inparaenum}
It is worth mentioning that the computational complexity of the former algorithm is moderate while the latter approach is \textit{not numerically tractable}. We hence use the asymptotic characterizations for these schemes~\cite{bereyhi2016statistical,bereyhi2016rsb}.

The performance is evaluated in terms of the average  \ac{mse} given by the average distortion when 
\begin{align}
\Delta\brc{x_n; \hat{x}_n } = \frac{1}{N} \abs{x_n - \hat{x}_n}^2.
\end{align}

The first experiment considers a codebook of size $S=1000$ and compares the classical approaches with an \ac{oas} scheme in which $M=80$. The results are shown in Fig.~\ref{fig:mse_vs_rate} for multiple compression rates. The \ac{oas} scheme uses Algorithm~\ref{alg:OAS} with a \textit{random} selection algorithm, i.e., the $K$ vectors are selected randomly in each subframe. As the result shows, the \ac{oas}-based algorithm performs close to LASSO in low compression rates, while at $R=5$, it even outperforms the MMSE bound.

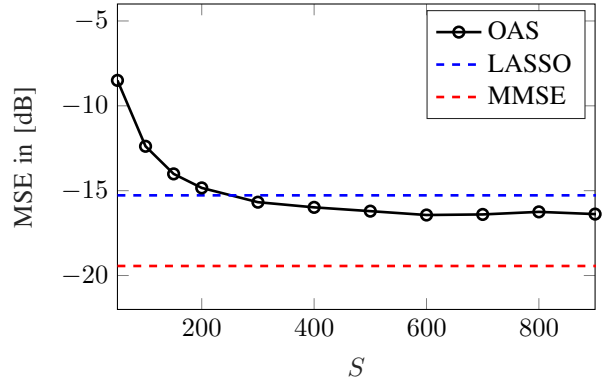
\begin{figure}[t]
%
%
\begin{tikzpicture}

\begin{axis}[%
width=2.5in,
height=1.6in,
at={(1.989in,1.234in)},
scale only axis,
xmin=50,
xmax=900,
xlabel style={font=\color{white!15!black}},
xlabel={$S$},
ymin=-22,
ymax=-4,
ylabel style={font=\color{white!15!black}},
ylabel={MSE in [dB]},
axis background/.style={fill=white},
legend style={legend cell align=left, align=left, draw=white!15!black}
]
\addplot [color=black, line width=1.0pt, mark=o, mark options={solid, black}]
  table[row sep=crcr]{%
50	-8.50108735117111\\
100	-12.3846970263169\\
150	-14.0116526807055\\
200	-14.8370003367178\\
300	-15.6790331303782\\
400	-15.9834453979235\\
500	-16.2070003948969\\
600	-16.4345105849813\\
700	-16.4010391524912\\
800	-16.2493192059775\\
900	-16.3804851212842\\
1000	-16.3729002352152\\
};
\addlegendentry{OAS}

\addplot [color=blue, dashed, line width=1.0pt]
  table[row sep=crcr]{%
50	-15.2756881199302\\
100	-15.2756881199302\\
150	-15.2756881199302\\
200	-15.2756881199302\\
300	-15.2756881199302\\
400	-15.2756881199302\\
500	-15.2756881199302\\
600	-15.2756881199302\\
700	-15.2756881199302\\
800	-15.2756881199302\\
900	-15.2756881199302\\
1000	-15.2756881199302\\
};
\addlegendentry{LASSO}

\addplot [color=red, dashed, line width=1.0pt]
  table[row sep=crcr]{%
50	-19.4385388842104\\
100	-19.4385388842104\\
150	-19.4385388842104\\
200	-19.4385388842104\\
300	-19.4385388842104\\
400	-19.4385388842104\\
500	-19.4385388842104\\
600	-19.4385388842104\\
700	-19.4385388842104\\
800	-19.4385388842104\\
900	-19.4385388842104\\
1000	-19.4385388842104\\
};
\addlegendentry{MMSE}

\end{axis}
\end{tikzpicture}%
	\caption{MSE against the codebook size. Here, the compression rate is set to $R=4$, number of subframes $M=60$ and $L = K/2 = 25$.}
	\label{fig:mse_vs_S}
\end{figure}

The behavior illustrated via Fig.~\ref{fig:mse_vs_rate} is consistent with the prior investigations on \ac{oas} with no predefined codebooks \cite{muller2018oversampled,muller2018randomoversampled,bereyhi2019oas}. In fact, those initial cases can be observed as a special case of \ac{oas} via a predefined codebook whose size goes to infinity. To investigate the performance degradation imposed by the codebook restriction, we further plot the \ac{mse} achieved via the \ac{oas}-based algorithm against the codebook size in Fig.~\ref{fig:mse_vs_S}. In this figure, we set $R=4$ and $M=60$. $L$ is further set to $L=K/2=25$. As the figure shows, the \ac{mse} drops fast and converges to the asymptotic value which specifies the \ac{oas} performance with no predefined codebooks.  

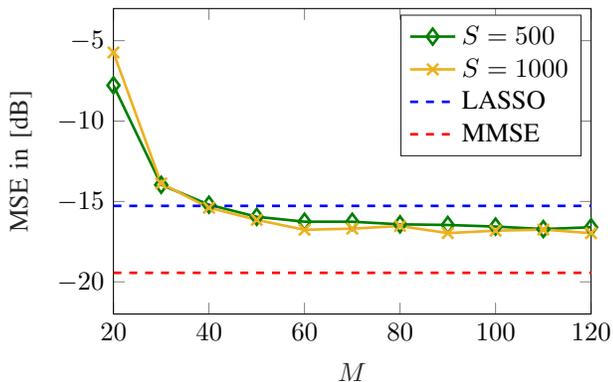
\begin{figure}[t]
%
%
\definecolor{mycolor1}{rgb}{0.00000,0.49804,0.00000}%
\definecolor{mycolor2}{rgb}{0.00000,0.44706,0.74118}%
\definecolor{mycolor3}{rgb}{0.92941,0.69412,0.12549}%
\begin{tikzpicture}

\begin{axis}[%
width=2.5in,
height=1.6in,
at={(1.989in,1.234in)},
scale only axis,
xmin=20,
xmax=120,
xlabel style={font=\color{white!15!black}},
xlabel={$M$},
ymin=-22,
ymax=-3,
ylabel style={font=\color{white!15!black}},
ylabel={MSE in [dB]},
axis background/.style={fill=white},
legend style={legend cell align=left, align=left, draw=white!15!black}
]



\addplot [color=mycolor1, line width=1.0pt, mark=diamond, mark options={mark size=3pt,solid, mycolor1}]
  table[row sep=crcr]{%
20	-7.77659290398459\\
30	-13.9650320902179\\
40	-15.200826654788\\
50	-15.9526552468812\\
60	-16.257504368707\\
70	-16.2604244252569\\
80	-16.421890876287\\
90	-16.4623940669881\\
100	-16.5618057704167\\
110	-16.7166227842952\\
120	-16.6041729817019\\
};
\addlegendentry{$S=500$}

\addplot [color=mycolor3, line width=1.0pt, mark=x, mark options={mark size = 3pt,solid, mycolor3}]
  table[row sep=crcr]{%
20	-5.73047505749962\\
30	-13.8772817614586\\
40	-15.389083351332\\
50	-16.1452268864302\\
60	-16.7612039661784\\
70	-16.6899535436441\\
80	-16.5272581019743\\
90	-16.9672307178716\\
100	-16.8187645685541\\
110	-16.7489341099981\\
120	-16.9722718628723\\
};
\addlegendentry{$S=1000$}

\addplot [color=blue, dashed, line width=1.0pt]
table[row sep=crcr]{%
	20	-15.2756881199302\\
	30	-15.2756881199302\\
	40	-15.2756881199302\\
	50	-15.2756881199302\\
	60	-15.2756881199302\\
	70	-15.2756881199302\\
	80	-15.2756881199302\\
	90	-15.2756881199302\\
	100	-15.2756881199302\\
	110	-15.2756881199302\\
	120	-15.2756881199302\\
};
\addlegendentry{LASSO}

\addplot [color=red, dashed, line width=1.0pt]
table[row sep=crcr]{%
	20	-19.4385388842104\\
	30	-19.4385388842104\\
	40	-19.4385388842104\\
	50	-19.4385388842104\\
	60	-19.4385388842104\\
	70	-19.4385388842104\\
	80	-19.4385388842104\\
	90	-19.4385388842104\\
	100	-19.4385388842104\\
	110	-19.4385388842104\\
	120	-19.4385388842104\\
};
\addlegendentry{MMSE}

\end{axis}
\end{tikzpicture}%
\caption{MSE against the number of subframes for different codebook sizes. Here, we set $R=4$ and $L = K/2 = 25$.}
\label{fig:mse_vs_M}
\end{figure}

Similar behavior is observed in Fig.~\ref{fig:mse_vs_M} in which the \ac{mse} is plotted against the number of subframes for two different codebook sizes. Here, we set $R=4$ and $L = K/2 = 25$. From the figure, it is observed that the \ac{oas}-algorithm converges to its asymptotic relatively fast, and further increase in the number of subframes improves the performance negligibly.

\subsection{Impact of Selection Algorithm}
Figs.~\ref{fig:mse_vs_rate}-\ref{fig:mse_vs_M} consider a \textit{random} selection algorithm. The performance of the \ac{oas} algorithm can be further improved by developing a more efficient selection algorithm. Such an algorithm can be developed by defining a concept of \textit{optimality}. In the sequel, we give an instance of such algorithms.

From derivations in Section~\ref{sec:derivations}, we know that Algorithm~\ref{alg:OAS} cancels out the residual samples in each subframe using the estimates given in the previous subframes; see \eqref{eq:approx}. This means that the algorithm relies on this postulation that estimated samples converge to good estimates as the algorithm evolves. Nevertheless, the approximated approach leads to unwanted interference terms in each subframe. 

The interference in the subframe $m$ is ideally suppressed, if the residual samples lie in the kernel of $\mW_m$, i.e. $\mW_m \bar{\bx}_m = 0$. However, this constraint is not necessarily fulfilled, since 
\begin{inparaenum}
	\item the residual samples are unknown, and
	\item $\mA_m$ is constructed from the codebook.
\end{inparaenum}
The first issue is addressed by following the approximation used in Algorithm~\ref{alg:OAS}, i.e., replacing~$\bar{\bx}_m$~with~$\tilde{\bx}_m$. For the second issue, an \textit{optimal} approach is to search the codebook, such that the sum-power of the interference terms is minimized, i.e., setting $\mA_m = \dbc{\ba_1\brc{m}, \ldots, \ba_K\brc{m} }^\trp$ where
\begin{align}
	\ba_1\brc{m}, \ldots, \ba_K\brc{m} = \argmin_{ \bu_1, \ldots, \bu_K \in \mac } \norm{ \mU \mE_m^\trp \tilde{\bx}_m }^2 \label{eq:optimal}
\end{align}
with $\mU = \dbc{\bu_1, \ldots, \bu_K }^\trp$.

The selection algorithm in \eqref{eq:optimal} deals with an exhaustive search whose number of choices exponentially grows by the codebook size. For practical scenarios with large codebooks, this selection algorithm is not numerically tractable. One can hence approximate the solution with a greedy algorithm. An example is Algorithm~\ref{alg1} which uses stepwise regression.
\begin{algorithm}[t]
	\caption{Adaptation Algorithm $\maa \brc{ \mac, K , \setF , \hat{\bx} }$}
	\label{alg1}
	\begin{algorithmic}[0]
		\In $K$, codebook $\mac$ and index set $\setF \subset \dbc{N}$.
		\Initiate Set $\ba_1 =\bc_{i_1}$ with $\bc_{i_1}$ being selected at random from $\mac$. Let $\setI_1 = \dbc{S} \backslash\set{i_1}$, $\mP=\sel \brc{\setF}$, $\mE = \sel\brc{ \dbc{N} \backslash \setF }$ and  $\tilde{\bx} = \mE \hat{\bx}$.\vspace*{.5mm}
		\For{$k \in \dbc{2:K}$}\\
		\begin{enumerate}
			\item Set $\mA_k = \dbc{ \baa_1, \ldots, \baa_{k-1}}$ and let
			\begin{align*}
				\bvv_k = \mP \mA_k \mA_k^\trp \mE^\trp \tilde{\bx}
			\end{align*} 
			\item For $\bu \in \setR^N$, let
			\begin{align*}
				f_k \brc{\bu} = \norm{  \brc{\bu^\trp \mE^\trp \tilde{\bx}} \mP \bu + \bvv_k }^2.
			\end{align*}
			\item Set $\ba_{k} = \bc_{i_k}$, where
			\begin{align*}
				i_k = \argmin_{i \in \setI_{k-1} } f_k \brc{\bc_i}.
			\end{align*}
			\item Update $\setI_k = \setI_{k-1} \backslash\set{i_k}$.
		\end{enumerate}
		\EndFor
	\end{algorithmic}
\end{algorithm}

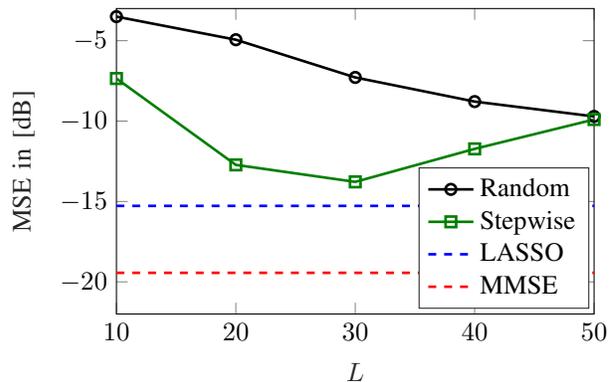
\begin{figure}[t]
%
%
\definecolor{mycolor1}{rgb}{0.00000,0.49804,0.00000}%
\begin{tikzpicture}

\begin{axis}[%
width=2.5in,
height=1.6in,
at={(1.989in,1.234in)},
scale only axis,
xmin=10,
xmax=50,
xlabel style={font=\color{white!15!black}},
xlabel={$L$},
ymin=-22,
ymax=-3,
ylabel style={font=\color{white!15!black}},
ylabel={MSE in [dB]},
axis background/.style={fill=white},
legend style={at={(.99,.48)},legend cell align=left, align=left, draw=white!15!black}
]

\addplot [color=black, line width=1.0pt, mark=o, mark options={solid, black}]
  table[row sep=crcr]{%
10	-3.50236821332384\\
20	-4.93947531947828\\
30	-7.28465771542162\\
40	-8.78683430032598\\
50	-9.71003990613216\\
};
\addlegendentry{Random}

\addplot [color=mycolor1, line width=1.0pt, mark=square, mark options={solid, mycolor1}]
table[row sep=crcr]{%
	10	-7.35231517008054\\
	20	-12.7242787255542\\
	30	-13.7813107482051\\
	40	-11.7281622260895\\
	50	-9.90123120493391\\
};
\addlegendentry{Stepwise}

\addplot [color=blue, dashed, line width=1.0pt]
  table[row sep=crcr]{%
10	-15.2756881199302\\
20	-15.2756881199302\\
30	-15.2756881199302\\
40	-15.2756881199302\\
50	-15.2756881199302\\
};
\addlegendentry{LASSO}

\addplot [color=red, dashed, line width=1.0pt]
  table[row sep=crcr]{%
10	-19.4385388842104\\
20	-19.4385388842104\\
30	-19.4385388842104\\
40	-19.4385388842104\\
50	-19.4385388842104\\
};
\addlegendentry{MMSE}

\end{axis}
\end{tikzpicture}%
	\caption{MSE versus $L$ for random and stepwise adaptations.  Here, the compression rate is set to $R=4$, codebook size $S=500$ and number of subframes is $M=20$.}
	\label{fig:mse_vs_L}
\end{figure}

Fig.~\ref{fig:mse_vs_L} compares the performance of Algorithm~\ref{alg1} with random selection. Here, Algorithm~\ref{alg:OAS} is run with both of the selection algorithms when $R=4$ and a codebook of size $S=500$ is available. The results are plotted for $M=20$ subframes. As the figure shows, the greedy algorithm outperforms the random selection which agrees with the intuition.

\section{Conclusions}
In presence of a predefined codebook, design of \ac{oas}-based algorithms deal with more challenges. This is due to the lower degrees of freedom provided by the setting. Although this restriction leads to degraded performance, the same behavior as the one, observed in the basic form of \ac{oas} framework, is reported. The performance degradation is further compensated by effective design of selection algorithms and higher number of subframes. These are feasible means in many applications in which \ac{oas} seems to be a good candidate.

\bibliography{ref}
\bibliographystyle{IEEEtran}
\end{document}